\documentstyle[prl,aps,multicol,epsf]{revtex}

\begin {document}
\def\inseps#1#2{\def\epsfsize##1##2{#2##1}\centerline{\epsfbox{#1}}}
\title { Cluster Dynamics for Randomly Frustrated Systems
with Finite Connectivity
}
\author{
N. Persky{\small $^1$},
I. Kanter{\small $^2$}, 
and S. Solomon{\small $^1$}\\
{\small $^1$}Racah Institute of Physics,
Hebrew University, Jerusalem 91904, Israel\\
{\small $^2$}Department of Physics,
Bar-Ilan University, Ramat-Gan 52900, Israel\\
{\small Emails: nathanp@vms.huji.ac.il, 
ido@kanter.ph.biu.ac.il,
sorin@vms.huji.ac.il.}\\
}
\maketitle
\begin{abstract} 
In simulations of some infinite range spin glass systems 
with finite connectivity, it is found that for
any resonable computational time, the saturated
energy per spin that is achieved by a cluster algorithm is lowered 
in comparison to that achieved by Metropolis dynamics.
The gap between the average energies 
obtained from these two dynamics 
is robust with respect to 
variations of the annealing schedule.
For some probability distribution of the interactions 
the ground state energy is calculated analytically
within the replica symmetry assumption
and is found to be saturated by a cluster algorithm. 
\end {abstract}
\vspace{-0.1in}
\begin{multicols}{2}
Many systems composed of a macroscopic number of interacting elements
share the property of being computationally 
difficult. By this we mean that
their relaxation time, as well as
the time scales related to the system investigation,
grow very fast with the size of the system \cite{par94}.
This concerns both the actual dynamics of the physical
systems as well as pseudo dynamics used in
computer
simulation \cite{par94}.
Such systems appear in many physical fields, from
statistical mechanics, to the study of quantum field
theories (for review see \cite{sorin1,Domany2}).

A typical 
example of such difficulties is the 
critical slowing down at
second order phase transitions.
This phenomenon is simply the divergence of the relaxation time
as the critical point is approached.
Consequently, the typical time needed to produce a large Boltzmann set of
decorrelated configurations
diverges and the standard local Monte Carlo
simulation methods become inefficient.

To solve such problems,
multiscale-cluster-type algorithms were devised, and
the entire subject of global collective dynamics
attracted considerable attention.
It is generally believed \cite{Domany2}
that for several classes of systems, multiscale methods
may overcome the slowing-down problems.
Moreover,
the multiscale algorithms are conceptually important insofar
as they encode the understanding of the
relevant large-scale physics.
In particular, these procedures
isolate the relevant degrees of freedom
and act directly on them,
in a manner consistent both with their effective macroscopic
dynamics, and with the basic interactions
that define the system at the microscopic scale (for a
review see Ref. \cite{sorin1}).
The cluster-multiscale methods have been applied successfully
to many fields in physics
(second-order phase
transitions, disordered systems,
quantum field theories, fermions in a gauge background,
quantum gravity, and more \cite{Domany2}), and in many cases a
dramatic acceleration of the numerical simulation was
achieved.

However, the general applicability of the
multi scale methods is in question.
The situation is particularly unclear
for models
with complex energy landscape.
Their physical properties are notoriously hard to investigate,
especially in the low-temperature phase.
Some of the most important families of systems
presenting such difficulties are the
randomly frustrated systems (RFS) and in particular
spin-glass (SG) systems.

The study of SG has attracted wide
research activity over the last
two decades (for review see \cite{BY,par3})
and their theoretical understanding
goes beyond its original scope
of understanding experimental
results of real physical systems \cite{BY}.
The progress of the statistical mechanics methods
related to SG contributed
to the understanding of a wide variety of other disordered systems.
One of the most promising directions is to apply the SG
knowledge to the study of hard optimization problems
belonging to the NP class \cite{par3}.
This relationship goes beyond a mere
analogy, and the task
of determining the optima of a problem can be rigorously mapped in to
finding the ground state (GS) of the analogous SG system.
A typical SG system presents a complex energy landscape
consisting of many local minima, separated by huge
barriers that
scale with the size of the system,
and lead to an infinite hierarchy
of exponentially divergent relaxation time scales \cite{sompo}.
Therefore, besides the problem of proper sampling in 
simulations at low temperatures, the system
has a tendency to get stuck in local minima
that prevent the measurement of equilibrium 
properties within a reasonable time.
Furthermore, the simulated annealing technique 
that prevents some systems from getting stuck
in local minima, failed to provide a complete solution in SG systems.

On the other hand, the {\em general} question of the existence of
efficient cluster algorithms (CA's) for
RFS is still an open problem even though
a number of successes have been
achieved for some particular systems.
For instance, Kandel et al. \cite{Domany} found an
efficient CA for a special case of
a fully frustrated system on a square lattice,
but with the lack of randomness.
In the special case of two-dimensions Ising SG,
in which its GS can be found 
in a polynomial time in contrast to the NP feature
of the general SG,
Swendsen and Wang \cite{SW2} developed their "replica" algorithm.
Nevertheless,
in general, the cluster algorithms are unable to identify
the important large-scale degrees of freedom
in RFS, and therefore
they show no improvement on simple local algorithms.

Hence, an efficient CA for RFS, if any, will enhance
our understanding of the low-temperature physics of
SG systems, and its consequences on related problems,
such as efficient heuristics algorithms for 
solving NP problems. 
In particular, finite connectivity models
at low temperatures
are directly connected to graph partitioning \cite{KS,Parisi}.
This is the problem of dividing a given graph into subgraphs,
with minimum connections between them.
Beside these applications, an efficient CA for SG may serve as a tool 
for the understanding of replica symmetry braking (RSB) and its properties.
Note that a quantitative fitting to Parisi's picture, and 
the existence of RSB in finite connectivity models, are 
still open and controversial questions.

In this letter we present a step towards understanding
the applicability and the limitation of CA for RFS.
We consider an Ising system described by the Hamiltonian
\begin{equation}
H=-{1\over2} \sum_{i\neq j}J_{ij}S_iS_j,
\label{m1}
\end{equation}
where $S_i=\pm1~ (i=1,...,N) $
and the probability distribution of the links is 
\begin{equation}
P(J_{ij})=(1-c/N)\delta(J_{ij})+(c/N)f(J_{ij}).
\label{m2}
\end{equation}
This model is known as a highly diluted system with
finite connectivity, since the probability 
for a spin with connectivity $k$
follows the Poisson distribution: $c^k\exp(-c)/k!$
with average connectivity $c$,
which is taken to be O(1), and 
$f(J_{ij}) $ is the distribution of the
surviving links after the dilution.
In the present work we would consider the following types
of unbiased [$f(J_{ij})=f(-J_{ij})$] distribution:
(a) Gaussian distribution, (b) $J_{ij}=\pm 1$, and
(c) a special case in which the links get 4 values:
$J_{ij}=\pm 1, J_{ij}=\pm \epsilon$.

This model with $ J_{ij}= \pm 1$ was systematically studied 
near the glass
transition temperature by Viana and Bary \cite{VB},
and at 
low temperatures by Kanter and Sompolinsky \cite{KS}
and by Mezard and Parisi \cite{Parisi}.
The self-consistent description of the low temperatures
is based on the probability distribution of
the local field defined by $ h_i \equiv T\tanh^{-1}<S_i>_T$.
Physically, this field is the first excitation, namely,
in the limit $T \rightarrow 0$, $|h_i|$ is the
minimum energy cost for flipping the $i$th spin from its GS
by the "best" reorganization of the system.
This local field is in truth an oxymoron, since it depends 
on {\bf \underline{global}} properties of the cluster,
the exchange field $\sum_j J_{ij}m_j$,
on the other hand, is truly a {\bf \underline{local}}
property depending on the local connectivity
(note that $|h_i| \leq |\sum_j J_{ij}m_j|$).
In a simple ferromagnetic case (that is J=1) as $T \rightarrow 0$
the distribution of the local fields reduces
to a discrete spectrum \cite{KS,Parisi}
\begin{equation}
P_l=(cP)^l \exp(-cP)/l!~l=0,...,\infty,
\label{pl}
\end{equation}
where $P$ is the fraction of the spins belonging to the 
macroscopic cluster. The quantities $P_l$, characterize
global properties, for instance, in the ferromagnetic
case $P_0$ is the fraction of spins belonging to
the finite clusters and ${P_1}$ is the fraction of spins that can be disconnected
from the infinite cluster by cutting only one link.

Many geometric properties of this system
are well understood\cite{KS,Parisi}.
In particular,
the system undergoes a percolation transition at $c=1$. 
The maximal cluster is of
$O(log N)$ for $c <1$,
$O(N^{2/3})$ at $c=1$, and of $O(N)$
for $c>1$ where its size is explicitly given by $P=1-P_0=1-\exp(-cP)$.

The topological structure of diluted models makes 
them good candidates
for employing CA's.
CA's usually
consist of two main steps. First,
blocks are constructed stochastically from many single 
elements where the links between them have been "frozen" according 
to their tendency to act coherently, and the other links deleted.
Second, updates are performed in which entire blocks flip rigidly,
in such a way that the Gibbs distribution is still fulfilled.
Accordingly,
the lattice splits into a set of clusters each formed by sites
that can be linked by a chain of frozen links.
For general SG systems this procedure fails because it freezes
the entire lattice into a single block.
However, for a highly dilute lattice the additional deletions
may be just enough to actually split the lattice into disjointed
blocks.
In this way,
the CA are effective in
locating and acting upon large regions
that interact weakly with the rest of the configuration.
In addition, the CA may have an efficient way of
pinning the frustration to weak links.
The global decisions made by the CA
on large-scale degrees of freedom are directly connected to the
structure of the field that presents global features.

Simulations on the model, Eqs. (1) and (2), and with a
Gaussian distribution for the links, $f(J) \propto \exp(-J^2)$,
were carried out using both
local (Metropolis)
and cluster (Wolff \cite{Wolff}) dynamics.
The simulations were carried out for various
connectivity values, and at temperatures below the 
glassy transition $T_c$.
The size of the system was between 1000 and 5000,
and the results were averaged over at least ten different samples.
A typical result is presented in Fig. 1,
and for all times monitored (up to 100 000 MCS per spin),
one can clearly see a gap
between the energies of the two dynamics.
Clearly, in the limit of an extremely long time, this gap should vanish,
but in practice it can be considered as two different
energy levels.
\begin{figure}
\narrowtext
\inseps{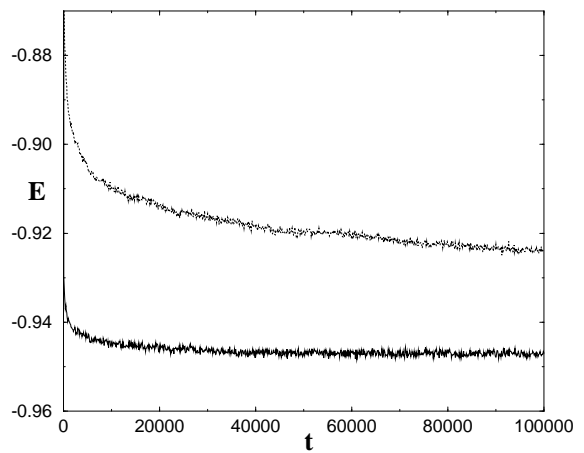}{0.45}
\caption{
$E(t)$ per spin for the Gaussian distribution
with average $|J|= 1$,
$c=2,N=5000$, and
$ T \sim 0.3T_c $. The solid and the dotted lines indicate
Wolff and Metropolis dynamics, respectively. The time 
scale is in MC steps per spin.
}
\end{figure}
An interesting question is whether this gap is robust even under
an annealed schedule
in which the temperature is gradually
lowered. This is done to avoid being caught in "false
minima", which is typical for low-temperature simulations, and
also enable the CA to act on clusters at various scales. 
Indeed, the results were improved for
both dynamics, but the gap still exists \cite{NSK}.
On one hand it is clear that in practice at low temperature,
the energy $E_c$ reached by the CA is lower
than that of the local algorithm $E_l$.
However, on the other hand, this improvement should be evaluated with 
respect to the true equilibrium energy, $E$.
Quantitatively, if $|E_c-E_l| \ll |E-E_l|$, then even after
the improvement by the CA, the system is far from equilibrium.
Hence, the calculation of the true GS is important 
in order to evaluate the improvement achieved by the CA.
Unfortunately, the analytical calculation of the GS energy
for the Gaussian distribution appears to be a very difficult task since 
the local field is a continuous variable\cite{NSK}.
We therefore need a model whose GS energy can be 
calculated analytically and 
where the difference between Metropolis 
and CA is enhanced.
We find these two features in what we shall call the 
{\em Weak-Strong} (WS) model.
In this model some of the links are much stronger (in their absolute 
value)
than others, and explicitly the link distribution
is given by
\vspace{-0.06 in}
\begin{eqnarray}
f(J)&=&{a\over2} [\delta(J-\epsilon)+\delta(J+\epsilon)] \nonumber \\
&+&{(1-a)\over2}[\delta(J-1)+\delta(J+1)]. \nonumber 
\end{eqnarray}

In the first set of runs we chose the connectivity $c$
and the fraction of the strong links $(1-a)$,
such that the density of the strong links by themselves is
below the percolation threshold, $c_S=c(1-a) < 1$. 
It is clear that in this situation all the strong links are
unfrustrated, and the frustration is located only on the weak links.
This framework simplifies the analytical treatment 
and only
the energy of the weak links, $E_W$, is considered.
In Fig. 2 one can see (up to our running time) a large steady 
difference ($ 35\%$) in the energy
between the local dynamics and that of the cluster dynamics.
This difference is due
to the fact that the local dynamics is totally stuck,
since the probability of flipping a cluster consisting of strong
links is practically zero for local algorithms.
On the other hand, the CA "knows" how to deal
with the strong link structures
by considering them as only "one degree of freedom"
for each cluster. In other words,
the CA is extremely efficient in the following
problem: "How to arrange the weak links
in the environment of strong links".
Note that the origin of this effect is the same as in the 
Gaussian case. In the WS case, however, it is amplified,
due to the special discrete link distribution.
In order to emphasize the CA features, we performed
measurements using simulated
annealing for both dynamics.
This was performed over a wide
range of temperatures,
enabling us to first arrange the strong links, and then
lowered the temperature to the scale of the weak links.
\begin{figure}
\narrowtext
\inseps{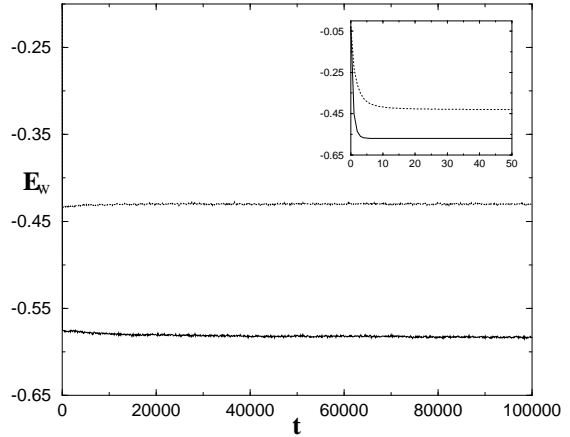}{0.44}
\caption{$E_{W}(t)$ per spin for the WS case, $c=2$, $a=0.7$, $N=5000$,
$ T=0.5 J_{W}$, the strong and the weak links was scaled to 
$100$ and $1$, respectively.
The solid and the dotted lines indicate
Wolff and Metropolis dynamics, respectively. 
Inset: The first 50 steps.
}
\end{figure}
Figure 3 demonstrates one of the examined schedules
from which one can conclude that the "gap" ($ 40\%$)
is robust (or even increases) with respect to the annealing schedule.
This provides strong evidence that the CA indeed do overcome the local
difficulties.
In order to put the CA improvement in the overall
picture of the true equilibrium, one 
must calculate the GS energy.
After some calculation, one can show that the GS energy
of the weak links
within the replica symmetry assumption is given by
\begin{equation}
E_W=-{1\over2}caP_0^2\epsilon+{1\over2}c(1-a)\sum_{k=0}^{\infty}(1-
4\sigma_k^2)\epsilon-\overline{h},
\label{ew}
\end{equation}
where $\sigma_k={P_0\over2}+\sum_{l=1}^kP_l$, $\overline{h}=\sum_{h=-\infty}^{\infty}{
|h|P(h)}$.
Note that the energy of the strong links, $E_S=-{1\over2}c(1-a)$,
is eliminated from Eq. (\ref{ew}).

The explicit value of $E_W$ depends on the local field,
$P(h)$, which in general is difficult to calculate.
Nevertheless, in the limit $c_S< 1 $ and
$ \epsilon \ll {1\over c} $,
which implies $h \ll 1$ (there are no "order 1"
local fields),
the following distribution of local fields can be assumed
\begin{equation}
P(h)=(1-Q)\delta(h)+\sum_{l\ne0}{P_l\delta(h-l \epsilon)} \label{ph1},
\end{equation}
where $P_l=P_{-l}$, and $Q=1-P_0$ the fraction of the frozen
(zero entropy at $T=0$)
spins.
The quantitative form of $P(h)$
can be determined from the following 
self-consistent equation:
\begin{eqnarray}
P(h)&=&e^{-cQ}\int_{-\infty}^{\infty}{{dy \over{2\pi}}\exp
\Bigl [
-iyh +{cQa\over2}(e^{iy\epsilon}+e^{-iy\epsilon})} \nonumber \\
&+&c(1-a)\sum_{l=1}^{\infty}{P_l(e^{iy l \epsilon}+
e^{-iy l\epsilon}) \Bigr ]}
\label{ph2}
\end{eqnarray}
In principle, one must solve infinitely many coupled non-linear
equations, obtained from the expansion of Eq. (\ref{ph2}).
However, assuming that $P_l$ decays with $l$,
we assume $P_l=0$ for $l>l'$, and then from the comparison of 
the expansion of Eq. (\ref{ph1}) and Eq. (\ref{ph2})
one can derive $l'$ nonlinear equations.
For the range of $c$ and $a$ in our simulations we take $l'=8$,
which gives a negligible error.
For comparison we calculated $P_l$ in the $J= \pm1$ case
in which we obtain
\begin{equation}
P_l=\exp(-cQ)I_{|l|}(cQ),
\end{equation}
where $I_l(x)$ is the modified Bessel function.
A graph of $P_l$ for the two cases
is presented in the inset of Fig. 3.

One should note that after scaling $\epsilon$
to $1$,
the $P_l$ for the two cases is very close.
However, the {\em exchange field} is a different story.
In the $J=\pm1$ case the exchange field of a spin is usually not
far from its local field (around the number of its neighbors),
but for the WS case the local field is $O({\epsilon})$ while the exchange field
may be $O(1)$.
The fact that $P_l$ is much smaller leads us to hope that the
dynamics that prefer a global "decision" are indeed
superior to the local one.

The type of the mean-field solution we derived
is known to be unstable \cite{JP}.
However, in Fig. 3 one can see that the analytical GS
obtained from Eqs. (\ref{ew}) and (\ref{ph2}),
is in very good agreement with the averaged GS energy
obtained by the CA.
\begin{figure}
\narrowtext
\inseps{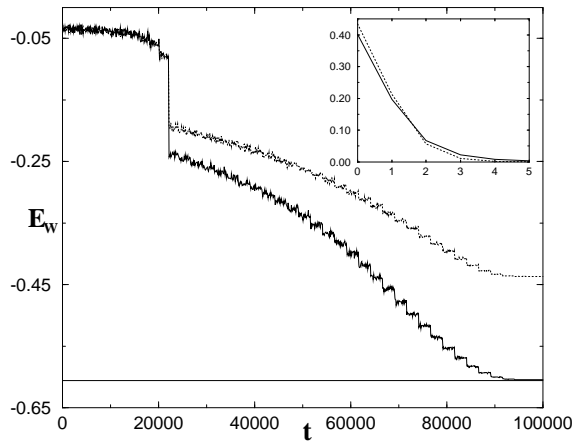}{0.45}
\caption{
 $E_{W}(t)$ per spin for the WS case, $c=2$ and $a=0.7$.
The annealing
schedule range is $T \in 123-0$, with $ \Delta T=10$ for $T>3$
and $\Delta T=0.1$ for $T<3$. 
The jump in $E_W$ at $T=3$ is observed since we are getting close to the scale
of the weak link and $ NE_W$ become extensive. The solid and the dotted lines 
indicate Wolff and Metropolis dynamics, respectively. 
The horizontal line denotes the analytical GS in Eq. (4).
Inset: $P_l$, for the WS case (solid), and $J=\pm 1$ case (dotted).
}
\end{figure}

The scope of our analytical calculation is 
for the $c_S<1$ case (below the percolation threshold for the strong links). 
However,
we also performed simulations for $c_S>1$, and the results 
for $c \in (2,5)$ show a similar picture.
Thus,
the superiority of CA over the local dynamics is shown clearly.
However, apriori, one could gain the impression that this was achieved only by 
some simple geometrical reduction of the system.
In order to check this point we performed another type of
simulations (which will be 
reported in detail elsewhere \cite {NSK}) in which the
algorithm explicitly reduced all the trees, the linear chains, and the
self-loops.
This was carried out in a manner that is proved to 
be energetically optimal.
Nevertheless, we show that
the CA advantage
goes beyond these simple reductions.
This result completes the picture of CA superiority,
within the scope of the present letter.
There are many questions 
that still remain open. First, what happens at 
high connectivity? It is clear that our
CA superiority decreases as $c$ grows.
One can investigate this point further, and classify the 
relevant windows of parameters.
Moreover, one can ask if it is possible to overcome
this limitation by a new type of CA,
namely, a CA that goes beyond the ability to act on blocks
having a considerably weak interaction with the rest of the system.
Second, an interesting question is whether CA can enter the RSB region.
Our results do not clarify this point, and
indeed our results are consistent with
the RS GS energy. However, the existence of 
RSB in the examined systems is still in question.

The research of I.K. and S.S. was
supported in part by the Israel Academy of Science,
the research of SS was
supported in part by the
Germany Israel Foundation (GIF).

\end{multicols}
\end{document}